\documentclass[11pt]{article}

\usepackage{amssymb,amsmath,graphicx}

\begin{document}

\begin{titlepage}

\begin{center}

\vspace{0.8cm}

{\Large Leptogenesis Scenarios via Non-Thermally Produced \\Right-handed Neutrino and Sneutrino\\
in Supersymmetric Seesaw Model}

\vspace{0.8cm}

\vspace{1.3cm}

{\bf Masato Senami}\footnote{
The author was in ICRR, Univ. of Tokyo, when this work started.
senami@me.kyoto-u.ac.jp}\\
{\it
Department of Micro Engineering,
Kyoto University,
Yoshidahonmachi, Kyoto,
606-8501, Japan
}\\
\vspace{0.3cm}
and \\
\vspace{0.3cm}
{\bf Tsutomu Takayama}\footnote{tstkym@icrr.u-tokyo.ac.jp}\\
{\it
Institute for Cosmic Ray Research,
 University of Tokyo, Kashiwa, Chiba 277-8582, Japan
}
 \\

\vspace{1cm}

\vspace{1.5cm}

\abstract{
We reconsidered leptogenesis scenario from right-handed (s)neutrino produced by the decay of inflaton.
Besides the well-investigated case that the neutrino decays instantaneously 
after the production, leptogenesis is possible if neutrino decays after it dominates the universe.
In the latter case, right-handed (s)neutrino can decay either while it is relativistic
or after it becomes non-relativistic.
Especially,  the first case has not been discussed seriously in literatures.
Resultant lepton asymmetry and constraints from the gravitino problem are studied in broad parameter region,
including all cases of this scenario.
It is also shown how this leptogenesis scenario depends on the parameters,
the inflaton decay rate (the reheating temperature), the right-handed neutrino mass,
the washout parameter, and the constraint from the gravitino problem.
Leptogenesis from relativistic neutrino decay is interesting 
because both thermal and non-thermal gravitino problems can be relaxed.
}

\end{center}
\end{titlepage}
\setcounter{footnote}{0}
\section{Introduction}

Leptogenesis \cite{Fukugita:1986hr} is 
an attractive scenario for explaining the baryon asymmetry of the universe,
$n_B/s \simeq (8.74\pm0.23)\times10^{-11}$ \cite{Komatsu:2008hk}, where $s$ is the entropy density.
This scenario is also appealing because it can be implemented within the seesaw model \cite{seesaw},
which explains the small non-zero neutrino mass.

In the simplest version of leptogenesis,
the lepton asymmetry is generated by the lepton number and $CP$ violating decay of 
thermally produced right-handed neutrino.
The lepton asymmetry is partially converted into the baryon asymmetry
via $(B+L)$ violating sphaleron processes \cite{sphalerons}.
In the minimal supersymmetric standard model (MSSM), 
the relation between the initial lepton asymmetry $n_L$ and 
the final baryon asymmetry $n_B$ is known to be
\cite{Harvey:1990qw}\footnote{
The ratio between $B-L$ and $B$ depends on details of the electroweak phase transition, 
the particle content and the mass spectrum \cite{sphaleron-detail}.
In this paper, we use eq.(\ref{eq:observed_baryon}) as a reference value,
which is obtained if all sparticles are sufficiently heavy and can be neglected.
}
\begin{eqnarray}
	\label{eq:observed_baryon}
	\frac{n_B}{s} = \frac{8}{23}\frac{n_L}{s}.
\end{eqnarray}
This scenario is known as thermal leptogenesis and investigated thoroughly in many literatures
\cite{thermal_leptogenesis}.

The result of detailed calculation in ref.\,\cite{Giudice:2003jh} shows that
$T_{\mathrm{reh}}\gtrsim M_N \gtrsim 10^9{\mathrm{GeV}}$ is required for successful leptogenesis.
Here, $T_{\mathrm{reh}}$ is the temperature of the universe at the beginning of 
radiation-dominant (RD) era, and $M_N$ is the mass of the lightest right-handed (s)neutrino.
$M_N > 10^9{\mathrm{GeV}}$ is required for large enough magnitude of $CP$ violation 
in decay processes of right-handed neutrino.
The condition $T_{\mathrm{reh}}\gtrsim M_N$ is required for thermal production of right-handed (s)neutrino in the thermal bath.

However, these conditions conflict with the constraint from the gravitino problem~\cite{gravitino_problem}.
In supersymmetric theories, $T_{\mathrm{reh}}$ 
is bounded for evading the overproduction of gravitino by thermal scattering
.
Thus, the bound on the reheating temperature constrains thermal leptogenesis scenario.
Both the abundance of gravitino produced by thermal scattering and
the effect of gravitino on the evolution of the universe depend on the gravitino mass, $m_{3/2}$.
Therefore, the bound from gravitino overproduction depends on $m_{3/2}$.
For a range $10^2{\mathrm{GeV}} < m_{3/2} < 10^4{\mathrm{GeV}}$, 
which is predicted in the gravity-mediated supersymmetry (SUSY) breaking, 
the constraint is $T_{\mathrm{reh}} \lesssim 10^{6-9}{\mathrm{GeV}}$.
In this case, thermal leptogenesis is severely constrained \cite{gravitino}.

One possible solution that reconciles leptogenesis scenarios with 
the gravitino problem is non-thermal leptogenesis scenarios.
The basic concept of these scenarios is very simple:
high reheating temperature $T_{\mathrm{reh}} > M_N$ is not required
if sufficient initial abundance of right-handed neutrino is generated without thermal scattering.

Many non-thermal leptogenesis scenarios have been proposed.
For example, 
leptogenesis from right-handed (s)neutrino generated by the inflaton decay
\cite{inflaton_decay, Lazarides:1996dv},
leptogenesis from the decay of right-handed sneutrino condensate 
\cite{Murayama:1993em,Hamaguchi:2001gw},
leptogenesis subsequent to sneutrino inflation \cite{sneutrino_inflation},
Affleck-Dine leptogenesis \cite{Affleck-Dine,Affleck-Dine_N,Senami:2007wz,Senami:2007up},
etc. (for example, \cite{etc_leptogenesis, etc_leptogenesis2}).

Here, we focus on leptogenesis from right-handed (s)neutrino produced by the inflaton decay.
In this scenario, inflaton is assumed to decay into right-handed (s)neutrinos,
with possible smaller branching ratio into MSSM particles, which are assumed to be thermalized immediately.
Then, the decay of right-handed (s)neutrino produces lepton asymmetry $n_L$.
Many models that can realize this scenario are proposed,
and in most of them the leptogenesis is considered only for the case that right-handed  (s)neutrino decays instantaneously 
after the production by inflaton decay.
Leptogenesis is also possible if neutrino decays after it dominates the universe.
In this case, gravitino problem can be avoided, since the universe is reheated
only after the decay of right-handed (s)neutrino.
Moreover, the right-handed (s)neutrino produced after inflaton decay is relativistic.
Hence, right-handed (s)neutrino can either decay while it is still relativistic
or after it becomes non-relativistic.
These cases have not been considered in most of recent literatures.
In ref.\,\cite{etc_leptogenesis}, this kind of non-thermal leptogenesis after 
neutrino dominance is discussed.
However, although they considered the era when right-handed (s)neutrino are relativistic,
they focused only on neutrino decay after non-relativistic neutrino dominance,
in addition to the case that right-handed  (s)neutrino decays instantaneously 
after the production.
As we discuss later, parameter region for leptogenesis from relativistic neutrino decay
is advantageous for relaxing both thermal and non-thermal gravitino problems \cite{gravitino_from_inflaton1,gravitino_from_inflaton-2}.
In this way, it is meaningful to reconsider all possible cases and show the distinction of three parameter regions.

In this work, we survey broad parameter region including all possibilities,
and  show conditions and parameter dependences of this scenario.
For simplicity, we assume that only one flavor $N_1$ is relevant to leptogenesis 
throughout this paper.
Therefore, free parameters in the neutrino sector are the lightest right-handed (s)neutrino mass $M_N$ of $N_1$ and 
the neutrino Yukawa coupling of $N_1$.
We also study the dependence on parameters of inflation,
the inflaton mass, the inflaton decay rate and the branching ratio into right-handed (s)neutrino of the inflaton
without specifying inflation models.

This paper is organized as follows.
As a preparation, in Section 2 we briefly review leptogenesis and the gravitino problem,
and introduce parameters that we will survey.
In Section 3, we consider the leptogenesis scenario from right-handed (s)neutrino 
produced by the inflaton decay.
We first discuss the estimation of the resultant lepton asymmetry and the constraint from
the gravitino problem, and then show the allowed parameter region.
The allowed parameter region is shown at last of this section.
We also discuss possible advantages of the parameter region which has not been considered in literatures.
We summarize our work in Section 4.

\section{Parameters of leptogenesis}

In this paper, we focus on leptogenesis scenarios from the decay of right-handed (s)neutrino
produced by non-thermal processes in the early universe.
In these scenarios, leptogenesis crucially depends on properties of the neutrino sector.

Our main purpose is to survey systematically the viability of non-thermal leptogenesis 
in broad region of seesaw model parameters without specifying inflation models.
Throughout this paper, we assume a hierarchical mass spectrum of right-handed neutrinos,
$M_{N_1}\ll M_{N_2} \ll M_{N_3}$.
In addition, in order to avoid complexity,
we restrict ourselves to the case that only the lightest right-handed (s)neutrino $N_1$ is relevant.
Hereafter, the subscript ``1'' is dropped.
Under this assumption, the superpotential relevant to the leptogenesis is given by 
\begin{eqnarray}
	\label{eq:superpotential}
	W = y_\nu N LH_u + \frac{1}{2}M_N NN + W_{\mathrm{MSSM}},
\end{eqnarray}
where $W_{\mathrm{MSSM}}$ is the superpotential of the MSSM sector.
Parameters of the neutrino sector relevant to leptogenesis scenarios are
the mass $M_N$ of (s)neutrino and the magnitude of the neutrino Yukawa coupling 
$(y_\nu y_\nu^\dagger)_{11}$.
The latter can be given by the washout parameter,
\begin{eqnarray}
	\label{eq:tilde_m}
	\tilde{m} \equiv \frac{(y_\nu y_\nu^\dagger)_{11} v_u^2}{M_N},
\end{eqnarray}
where $v_u$ is the vev of the up-type Higgs, $v_u=\sin\beta \times 174{\mathrm{GeV}}$.
Hereafter, we use $\tilde{m}$ as a parameter which indicates the magnitude of the neutrino Yukawa coupling.

Lepton asymmetry is generated by the out-of-equilibrium $CP$ violating decay of right-handed 
(s)neutrino.
The decay rate of (s)neutrino is calculated as 
\begin{eqnarray}
	\label{eq:Gamma_N}
	\Gamma_N = \frac{(y_\nu y_\nu^\dagger)_{11} M_N}{4\pi} = \frac{\tilde{m} M_N^2}{4\pi v_u^2},
\end{eqnarray}
in the leading order.
The temperature of the thermal bath at the decay of (s)neutrino is given by
\begin{eqnarray}
	\label{eq:T_N}
	T_N = \left( \frac{90}{\pi^2 g_*} \right)^\frac{1}{4}\sqrt{\Gamma_N M_{\mathrm{Pl}}}
	\simeq 0.12M_N\times\left( \frac{\tilde{m}}{10^{-5}{\mathrm{eV}}} \right)^\frac{1}{2},
\end{eqnarray}
regardless of whether the right-handed (s)neutrino dominate the universe at the decay.
Here, $M_{\mathrm{Pl}} = 2.4\times10^{18}{\mathrm{GeV}}$ is the reduced Planck mass.
We also assume $\sin\beta\simeq 1$ and $g_*\simeq200$ and omit dependences on these parameters.

We use the following expression of $CP$-asymmetry in the decay of $N$ \cite{Buchmuller:1998zf}:
\begin{eqnarray}
	\label{eq:epsilon}
	\epsilon &\equiv& 
	\frac{\Gamma(N\to H_u  l) -\Gamma (N \to \bar{H}_u\bar{l})}
	{\Gamma(N\to H_u  l) +\Gamma (N \to \bar{H}_u\bar{l})} \nonumber \\
	&\simeq& \frac{3}{8\pi} \frac{M_N}{v_u^2}m_{\nu_3} \delta_{\mathrm{eff}}\nonumber \\
	&\simeq& 2\times10^{-10} \times \left( \frac{M_N}{10^6{\mathrm{GeV}}} \right) 
	\left( \frac{m_{\nu_3}}{0.05{\mathrm{eV}}} \right)\delta_{\mathrm{eff}},
\end{eqnarray}
where $\delta_{\mathrm{eff}}$ is the effective $CP$ violating phase.
In the second line, $M_{N_1}\ll M_{N_2} \ll M_{N_3}$ and
a normal hierarchical mass spectrum of light neutrinos are assumed.
From results of atmospheric neutrino oscillation experiments,
the heaviest light neutrino mass is suggested to be $m_{\nu_3}\approx\sqrt{\Delta m_{\mathrm{atm}}^2}
\approx0.05{\mathrm{eV}}$ \cite{atmospheric}.
This $\epsilon$ is the expectation value of lepton number generated by the decay of one $N$ 
or $\tilde{N}$ particle.

We will concentrate on the case that the decay process become effective 
and lepton asymmetry is generated for $T_N<M_N$.
In this case, the washout of lepton asymmetry by inverse decay processes can be safely neglected.
Conversely, if the decay and inverse decay processes become effective for $T_N>M_N$, 
partial or complete equilibrium is maintained after the production of lepton asymmetry.
Thus, the amount of the lepton asymmetry is reduced by washout effects.
Successful leptogenesis is still possible even for $T_N>M_N$, if the washout is not so strong.
However, in this case, Boltzmann equations must be solved to estimate the 
amount of resultant lepton asymmetry.
This goes beyond the scope of our work.

For $T_N<M_N$, we can also safely neglect the washout of lepton asymmetry by various scattering processes.
Effective processes are relevant to the top Yukawa coupling, like $l+\bar{q}_3 \to N+\bar{t}$.
The amount of (s)leptons and (s)quarks that can produce heavy right-handed (s)neutrino
by these scattering processes are Boltzmann-suppressed for $T<M_N$.
Thus, these scattering processes are not effective.

It should be noted that these diagrams give early thermalization of right-handed (s)neutrino,
e.g.\,$N+q_3\to l+t$.
If these processes are effective, the abundance of non-thermally produced right-handed (s)neutrino
is decreased.
We will discuss later the possibility of early thermalization.

Next, let us consider inflation models.
In this work, we do not specify inflation models.
We assume that the thermal history after the inflation can be determined 
by the decay rate of inflaton $\Gamma_\phi$.
If the universe becomes RD after the decay of the inflaton,
the temperature of the radiation at the beginning of the RD universe $T_{\mathrm{reh}}$ is estimated to be 
$T_R$, which is given by $\Gamma_\phi$, 
\begin{eqnarray}
	\label{eq:T_R}
	T_R = \left( \frac{90}{\pi^2 g_*} \right)^\frac{1}{4}\sqrt{\Gamma_\phi M_{\mathrm{Pl}}}.
\end{eqnarray}
It should be noted that, if right-handed (s)neutrino dominate the universe before they decay,
$T_{\mathrm{reh}}$ is not $T_R$ but $T_N$.
In this case, the parameter $T_R$ does not stand for the temperature of the thermal bath after inflaton decay
since inflaton is assumed to decay mainly into right-handed (s)neutrino,
and the energy of thermal bath accounts for only a small fraction of the total energy of the universe.

Finally, we review the constraint on $T_{\mathrm{reh}}$ from the gravitino problem.
As mentioned in the introduction,
the gravitino problem is one of our principal motivation to consider non-thermal leptogenesis.
If there is no dilution process after the universe becomes RD, the amount of gravitino produced by thermal processes depends on 
$T_{\mathrm{reh}}$.
Thus, the dependence on $T_{\mathrm{reh}}$ is easily estimated by the following discussion.
By thermal processes, gravitino is produced in the amount of
\begin{eqnarray}
	\label{eq:gravitino}
	\left.\frac{n_{3/2}}{s}\right|_{H=\Gamma_\phi}\sim 
	\frac{\langle \sigma_g v\rangle n^2}{H}\frac{1}{s}
	\sim \frac{T_{\mathrm{reh}}}{M_{\mathrm{Pl}}},
\end{eqnarray}
where $n$ is the total number density of all relevant species in the thermal bath,
and $\sigma_g$ is a typical cross section of gravitino producing processes.
More accurate calculation can be found in ref.\,\cite{gravitino,Bolz:2000fu}.
Hereafter we refer to the result,
\begin{eqnarray}
	\label{eq:gravitino-refer}
	\left.\frac{n_{3/2}}{s}\right|_{H=\Gamma_\phi}\simeq 1.9\times10^{-12}
	\times\left( \frac{T_{\mathrm{reh}}}{10^{10}{\mathrm{GeV}}}\right).
\end{eqnarray}
The abundance of gravitino is constrained by the cosmology.
For unstable gravitino, the abundance is constrained for successful 
big-bang nucleosynthesis (BBN),
because decay products of gravitino alter abundances of light elements.
This constraint is thoroughly investigated in ref.\,\cite{gravitino}.
For $10^2{\mathrm{GeV}} < m_{3/2} < 10^4{\mathrm{GeV}}$,
which is predicted by the gravity-mediated SUSY breaking,
the constraint on $T_{\mathrm{reh}}$ is $T_{\mathrm{reh}} < 10^{6-7}{\mathrm{GeV}}$ if gravitino decays mainly in hadronic channel
and $T_{\mathrm{reh}} < 10^{6-9}{\mathrm{GeV}}$ if it decays mainly in non-hadronic channel.
Thermal leptogenesis is difficult under these constraints on $T_{\mathrm{reh}}$.
Hence, we focus on this bound on $T_{\mathrm{reh}}$,
and consider non-thermal leptogenesis as a candidate that 
can solve the conflict of leptogenesis with the gravitino problem.
For later convenience, we introduce a parameter $T_g$, 
which is defined as the maximum allowed reheating temperature for avoiding the gravitino problem.

We comment on other way out of this problem.
For example, $T_g>10^9{\mathrm{GeV}}$ is possible for $m_{3/2}\gtrsim10^4{\mathrm{GeV}}$,
which can be realized in anomaly-mediated SUSY breaking \cite{AMSB}.
On the other hand, the gravitino problem as well as other cosmological problems
can be avoided if $m_{3/2} < 16{\mathrm{eV}}$ \cite{Viel:2005qj},
which can be realized in gauge-mediated SUSY breaking \cite{GMSB1,GMSB2}.

In order to parameterize inflation models, we introduce
the mass of the inflaton\footnote{
As a reference, $m_\phi$ in several major examples of inflation models are summarized as follows.
Typically, $m_\phi$ in chaotic inflation with quadratic potential of inflaton 
is required to be $m_\phi\sim10^{13}{\mathrm{GeV}}$ by cosmological perturbations.
Successful cosmology requires $10^{10}{\mathrm{GeV}} < m_\phi < 10^{15}{\mathrm{GeV}}$ 
in $F$-term hybrid inflation models \cite{hybrid}, 
and  $m_\phi \sim 10^{10}{\mathrm{GeV}}$ for $m_{3/2}\sim 0.1-100{\mathrm{TeV}}$
in single-field new inflation models within supergravity \cite{Ibe:2006fs}.
} $m_\phi$ and the branching ratio into right-handed (s)neutrino $B_N$.
We will study the dependence of leptogenesis scenario on following parameters:
$\Gamma_\phi,~M_N,~\tilde{m},~m_\phi$ and $B_N$.

\section{Leptogenesis from right-handed (s)neutrino produced by the decay of inflaton}

One  of interesting possibilities is that heavy right-handed (s)neutrino are generated
by the decay of inflaton $\phi$.
Examples of inflation models and interactions in superpotential or K\"ahler potential 
that give inflaton decay dominantly into right-handed neutrino are found in refs.\,\cite{inflaton_decay, Lazarides:1996dv}.

In this section, we investigate the conservative bound on this scenario,
following carefully  the evolution of the universe after the inflaton decay.
We work  on as general framework as possible.
We treat the decay rate of the inflaton $\Gamma_\phi$ as a free parameter.
Here we assume that the inflaton can dominantly decay into right-handed (s)neutrino,
without specifying inflation models and interactions between the inflaton sector and the neutrino sector.
Hence, we assume that the branching ratio into right-handed (s)neutrino is $B_N\simeq 1$,
and that into the thermal bath consists of MSSM particles $B_{\mathrm{rad}}$ is negligibly small.
We later comment also on the case that $B_{\mathrm{rad}}$ is not negligible.
The right-handed neutrino mass $M_N$ and the neutrino Yukawa coupling are also 
treated as parameters.
Note that, for simplicity, we consider only 2-body decay processes $\phi\to N+N$ or 
$\phi\to \tilde{N}+\tilde{N}$.
These processes are allowed only if $M_N < m_\phi/2$.
We also assume that only $N_1$ is relevant.
This assumption is realized,
for example, if decay processes of the inflaton into $N_2$ and $N_3$ are kinematically forbidden\footnote{
	Almost the same discussion can be applied to the cases that the inflaton decays
	dominantly into $N_2$ or $N_3$, and branching ratios into lighter generations of right-handed neutrino 
	are negligibly small.
	In order to avoid the washout, the temperature of the thermal bath after (s)neutrino decay
	must be lower than the mass of the lightest right-handed (s)neutrino $N_1$.
}.

This scenario can be classified into three cases.
\begin{enumerate}
\renewcommand{\labelenumi}{(\roman{enumi})}
\item The decay of (s)neutrino is faster than that of the inflaton.
\item The decay of right-handed (s)neutrino is slower than that of the inflaton,
and right-handed (s)neutrino decays after it becomes non-relativistic.
\item The decay of right-handed (s)neutrino is slower than that of the inflaton,
and right-handed (s)neutrino decays during it is relativistic.
\end{enumerate}
In the case (i), right-handed (s)neutrino decays instantaneously after their production by the inflaton decay.
Latter two cases are interesting, since thermal production of gravitino takes place
only after right-handed (s)neutrino decay,
and the gravitino problem can be ameliorated.
As a bonus, in the case (ii), other unwanted relics produced at the decay of the inflaton
can be diluted by the entropy production from decaying right-handed (s)neutrino.
On the other hand, in the case (iii), since right-handed (s)neutrino is relativistic
at their decay, no dilution process exists,
compared with the case that the inflaton decays dominantly into the radiation.
These cases have not been considered in most of recent literatures.
In ref.\,\cite{etc_leptogenesis}, although the fact is taken into account that right-handed (s)neutrino is relativistic
after the production,
case (iii) is not seriously investigated.

In the next subsection, we discuss these three cases in order
and clarify the distinction of three cases.
We investigate the amount of lepton asymmetry, constraints from the gravitino problem,
and the condition for avoiding the washout.
We show that the successful leptogenesis is possible in all cases.
Finally, we summarize our results and discuss on possible advantages of the case (iii).

\subsection{Lepton asymmetry and gravitino constraint}
First, we consider the case (i).
Here, the fact should be taken into consideration that energy of (s)neutrino is $m_\phi/2$
at the  production\footnote{
	If right-handed neutrinos are produced by $n$-body decay with ($n>2$),
	momentum of neutrino is distributed around $m_\phi/n$.
	Our discussion is easily extended to these cases.
}.
This results in suppression of the decay rate of the (s)neutrino 
by a factor of $M_N/E_N$, where $E_N$ is the energy of right-handed (s)neutrino.
Therefore, this case is realized under the condition, $\Gamma_N\times(2M_N/m_\phi) > \Gamma_\phi$,
which is reduced to
\begin{eqnarray}
	\label{eq:cond-Gamma_N>Gamma_phi}
	M_N > 3.3\times10^{10}{\mathrm{GeV}} \times
	\left( \frac{T_R}{10^9{\mathrm{GeV}}} \right)^\frac{2}{3}
	\left( \frac{\tilde{m}}{10^{-5}{\mathrm{eV}}} \right)^{-\frac{1}{3}}
	\left( \frac{m_\phi}{10^{12}{\mathrm{GeV}}} \right)^\frac{1}{3}.
\end{eqnarray}
The estimation of the lepton asymmetry is straightforward.
Right-handed (s)neutrino decay instantaneously after their production
and the almost all energy of inflaton is released to the radiation.
Hence, the reheating temperature is estimated to be $T_R$.
The number density of inflaton is estimated by dividing the total energy by the mass of the inflaton.
Since $2B_N$ (s)neutrino is produced by decay of one inflaton, the final lepton asymmetry is
\begin{eqnarray}
	\label{eq:L_IDLG1}
	\frac{n_L}{s} &=& \frac{3}{2} \epsilon B_N \frac{T_R}{m_\phi} \nonumber \\
	&=& 3\times 10^{-10} \times B_N 
	\left( \frac{T_R}{10^6{\mathrm{GeV}}} \right)
	\left( \frac{M_N}{m_\phi} \right)
	\delta_{\mathrm{eff}}.
\end{eqnarray}
According to eq.\,(\ref{eq:L_IDLG1}), successful leptogenesis requires $T_R > 10^6{\mathrm{GeV}}$.
The washout of lepton asymmetry due to inverse decay processes can be neglected if $T_R<M_N$.
Thermalization of right-handed (s)neutrino before their decay can be neglected,
because they decay instantaneously after their production.
Since scattering processes like $N+t \to l+q_3$ are slower than the inverse decay for $T<M_N$,
effects of these processes on the lepton asymmetry are safely neglected.

Because the RD universe is realized after the decay of the inflaton in this scenario,
the gravitino problem requires low reheating temperature $T_R<T_g$.
It should be noted that, as $T_g$ decreases,
the allowed range of $M_N$ is restricted to a narrow range close to $m_\phi/2$.

Let us proceed to the case (ii).
The condition for the case (ii) is estimated as follows.
Right-handed (s)neutrino have momentum $m_\phi/2$ just after the production,
and the redshift of the momentum is $p\propto H^{1/2}$.
Therefore, right-handed (s)neutrino become non-relativistic at
\begin{eqnarray}
	\label{eq:H_NR}
	H=H_{\mathrm{NR}} \equiv \Gamma_\phi\left( \frac{2M_N}{m_\phi} \right)^2.
\end{eqnarray}
Right-handed (s)neutrino decay after they  become non-relativistic if $H_{\mathrm{NR}} > \Gamma_N$.

Since $T_{\mathrm{reh}}$ is estimated to be $T_N$,
the number density of the right-handed (s)neutrino at the decay is estimated simply 
by dividing $\rho_N \sim \rho_{\mathrm{tot}}\sim T_N^4$ by $M_N$,
where $T_N$ is given in eq.\,(\ref{eq:T_N}).
The number density and the momentum distribution of right-handed (s)neutrino 
are not changed by scattering, and are simply redshifted by the expansion of the universe.
Thus, the resultant lepton asymmetry is
\begin{eqnarray}
	\label{eq:L_IDLG2}
	\frac{n_L}{s} = \frac{3}{4}\epsilon\frac{T_N}{M_N}
	\simeq 1.5\times10^{-10} \times
	\left( \frac{T_N}{10^6{\mathrm{GeV}}} \right)
	\delta_{\mathrm{eff}},
\end{eqnarray}
which depends only on properties of neutrino.
The possibility of early thermalization of right-handed (s)neutrino 
by scattering processes like $N + t \to l + q_3$ can be neglected, 
because, due to small $B_{\mathrm{rad}}$, number density of MSSM particles in the initial state is small\footnote{
	When we consider larger $B_{\mathrm{rad}}\sim\mathcal{O}(1)$, early thermalization
	is not negligible if the neutrino Yukawa coupling is large.
	The most stringent constraint arises in the case that right-handed (s)neutrino already 
	becomes non-relativistic at $T=M_N$ and the right-handed (s)neutrino still does not dominate the universe.
}.
In processes like $N+N\to l+l$, the number density of initial particles are not small.
However, in the parameter region we are interested in,
these processes are also negligible because of the small neutrino Yukawa coupling.

Washout processes are safely avoided if $T_N<M_N$,
which is satisfied for
\begin{eqnarray}
	\label{eq:washout}
	\tilde{m}<7.3\times10^{-4}{\mathrm{eV}}.
\end{eqnarray}
Hereafter, we will focus on this bound on $\tilde{m}$.
It is expected that even for slightly larger $\tilde{m}$ successful leptogenesis may be realized,
although the viable parameter region may be restricted due to partial washout of the lepton asymmetry.
Unfortunately, the estimation of this bound requires to solve Boltzmann equations.
In ref.\,\cite{HahnWoernle:2008pq}, Boltzmann equations of this system is solved,
although right-handed (s)neutrinos are always assumed to be non-relativistic.
In our work, we included relativistic (s)neutrino,
while we leave detailed numerical study as a future work 
since this estimation is too detailed for our purpose. 

Constraints on this scenario are estimated as follows.
Since the universe is completely RD after the decay of the right-handed (s)neutrino,
the temperature after the decay of right-handed (s)neutrino $T_N$ 
is constrained as $T_N<T_g$.
Using eq.\,(\ref{eq:T_N}), this is given by 
\begin{eqnarray}
	\label{eq:cond-T_N<gravitino}
	M_N < 8.6\times10^{6}{\mathrm{GeV}}\times
	\left( \frac{\tilde{m}}{10^{-5}{\mathrm{eV}}} \right)^{-\frac{1}{2}}
	\left( \frac{T_g}{10^6{\mathrm{GeV}}} \right)
	.
\end{eqnarray}

Let us comment on the case if $B_{\mathrm{rad}}$ is not negligibly small.
In this case, gravitino produced by thermal processes after the inflaton decay
is also problematic.
In order to use the result of detailed estimations eq.\,(\ref{eq:gravitino-refer}),
we estimate the ratio between abundances of gravitino after the reheating in two different scenarios:
with and without extra entropy production by decay of right-handed (s)neutrino after the inflaton decay.
Here we consider that the case (i) is included in the latter scenario,
where the abundance of gravitino is given by eq.\,(\ref{eq:gravitino-refer}).
After the inflaton decay,  gravitino is produced by thermal processes as,
\begin{eqnarray}
	\label{eq:gravitino_in_thermal_bath}
	n_{3/2} 
	\sim \frac{\langle \sigma_g v \rangle n_{\mathrm{MSSM}}^2}{\Gamma_\phi}
	\sim \frac{T_{\mathrm{rad}}^6}{M_{\mathrm{Pl}}^2} \frac{1}{\Gamma_\phi}
	\sim B_{\mathrm{rad}}^\frac{3}{2}
	\frac{T_R^4}{M_{\mathrm{Pl}}}.
\end{eqnarray}
Thus, taking non-negligible $B_{\mathrm{rad}}$ into account, the amount of gravitino produced 
after the decay of the inflaton is given with a factor $B_{\mathrm{rad}}^{3/2}$.
As long as (s)neutrino are relativistic,
the expansion of the universe is $H\propto a^{-2}$, where $a$ is the scale factor, 
until the right-handed (s)neutrino become non-relativistic.
Following Hubble expansion of the universe,
we can estimate the abundance of gravitino after the decay of right-handed (s)neutrino as
\begin{eqnarray}
	\label{eq:gravitino_in_case2}
	\left.\frac{n_{3/2}}{s}\right|_{H=\Gamma_N}
	= 1.9\times10^{-12}
	\times\left( \frac{T_{\mathrm{R}}}{10^{10}{\mathrm{GeV}}}\right)
	\times B_{\mathrm{rad}}^\frac{3}{2}\left( \frac{\Gamma_N}{\Gamma_\phi} \right)^\frac{1}{2}
	\left( \frac{m_\phi}{2M_N} \right).
\end{eqnarray}
Combining this result and the constraint on the abundance of gravitino given in Section\,2,
the gravitino problem requires $B_{\mathrm{rad}}^{3/2}T_N(m_\phi/2M_N)<T_g$.
This condition can be given by
\begin{eqnarray}
	\label{eq:const-gravitino_in_case2}
	B_{\mathrm{rad}}^\frac{3}{2}m_\phi < 1.7\times10^7{\mathrm{GeV}} \times 
	\left( \frac{\tilde{m}}{10^{-5}{\mathrm{eV}}} \right)^{-\frac{1}{2}}
	\left( \frac{T_g}{10^6{\mathrm{GeV}}} \right).
\end{eqnarray}
While this constraint can be easily satisfied by assuming small $B_{\mathrm{rad}}$,
this may constrain interaction between the inflaton sector and the other particles
for high-scale inflation models (see fig.\,\ref{fig:IDLG-2}).

Note that other unwanted relics produced at the inflaton decay,
such as gravitino produced by direct decay from the inflaton \cite{gravitino_from_inflaton-2},
are also diluted by the dilution factor
\begin{eqnarray}
	\label{eq:dilution}
	\Delta &=& \left( \frac{\Gamma_N}{\Gamma_\phi} \right)^\frac{1}{2}
	\left( \frac{m_\phi}{2M_N} \right) \nonumber \\
	&=& \left( \frac{T_N}{T_R} \right)
	\left( \frac{m_\phi}{2M_N} \right) \nonumber \\
	&=& 0.06\times\left( \frac{\tilde{m}}{10^{-5}{\mathrm{eV}}} \right)^\frac{1}{2}
	\left( \frac{T_R}{10^{12}{\mathrm{GeV}}} \right)^{-1}
	\left( \frac{m_\phi}{10^{12}{\mathrm{GeV}}} \right).
\end{eqnarray}
Where, dilution factor $\Delta$ is the ratio between abundances of gravitino in two different scenarios,
with and without entropy production by decay of right-handed (s)neutrino.

Let us consider the rest one, the case (iii).
Since the decay rate of relativistic right-handed (s)neutrino is suppressed,
the Hubble parameter at the decay $H_{\mathrm{dec}}$ is estimated by solving
\begin{eqnarray}
	\label{eq:delayed-decay}
	H_{\mathrm{dec}} = \displaystyle \Gamma_N\times\frac{M_N}{\frac{m_\phi}{2}
	\left( \frac{H_{\mathrm{dec}}}{\Gamma_\phi} \right)^\frac{1}{2}}.
\end{eqnarray}
Now $T_{\mathrm{reh}}$ is estimated as $T_{\mathrm{dec}}$,
which is given by
\begin{eqnarray}
	\label{eq:T_dec}
	T_{\mathrm{dec}} = \left( \frac{90}{\pi^2 g_*} \right)^\frac{1}{4}
	\sqrt{H_{\mathrm{dec}}
	M_{\mathrm{Pl}}}.
\end{eqnarray}
Like the case (ii), early thermalization of right-handed (s)neutrino is negligible.
The number density of (s)neutrino at the decay is estimated by taking the expansion of the universe
into account.
The final lepton asymmetry is estimated as
\begin{eqnarray}
	\label{eq:L_IDLG3}
	\frac{n_L}{s}\simeq \frac{3}{4}\epsilon \times2B_N \frac{T_R^4}{m_\phi}
	\left( \frac{H_{\mathrm{dec}}}{\Gamma_\phi} \right)^\frac{3}{2}\frac{1}{T_{\mathrm{dec}}^3}
	= \frac{3}{2}\epsilon B_N \frac{T_R}{m_\phi} .
\end{eqnarray}
Note that $T_R$ defined in eq.\,(\ref{eq:T_R}) is not $T_{\mathrm{reh}}$.
Although the universe is dominated by the right-handed (s)neutrino, 
this result is the same as eq.\,(\ref{eq:L_IDLG1}).
Since the expansion of the universe is similar to that of the RD
universe, the entropy-to-(s)neutrino number density ratio does not change after the inflaton decay.
This also means that gravitino produced at the inflaton decay cannot be diluted.
Finally, since the temperature of the thermal bath at the decay of the right-handed (s)neutrino is 
$T_{\mathrm{dec}}$, the condition for negligible washout is $T_{\mathrm{dec}} < M_N$, or,
equivalently,
\begin{eqnarray}
	\label{eq:thermalize-case3}
	\tilde{m} < 3.7\times10^{-2}{\mathrm{eV}}\times
	\left(\frac{m_\phi}{10^{12}{\mathrm{GeV}}} \right)
	\left(\frac{T_R}{10^{10}{\mathrm{GeV}}} \right)^{-1}.
\end{eqnarray}
In the following discussion, we take the more stringent bound eq.\,(\ref{eq:washout}),
in order to safely neglect washout processes.

The gravitino problem puts constraint on the temperature of the thermal bath after the (s)neutrino decay,
$T_{\mathrm{dec}}<T_g$, or, equivalently, 
\begin{eqnarray}
	\label{eq:T_dec<gravitino}
	M_N < 3.3\times10^7{\mathrm{GeV}}\times
	\left( \frac{T_R}{10^9{\mathrm{GeV}}} \right)^{-\frac{1}{3}}
	\left( \frac{\tilde{m}}{10^{-5}{\mathrm{eV}}} \right)^{-\frac{1}{3}}
	\left( \frac{m_\phi}{10^{12}{\mathrm{GeV}}} \right)^\frac{1}{3}
	\left( \frac{T_g}{10^6{\mathrm{GeV}}} \right)
	.
\end{eqnarray}
If $B_{\mathrm{rad}}$ is not negligibly small, gravitino produced after the inflaton decay should be 
taken into the consideration.
The abundance of this gravitino have already been estimated
in eq.\,(\ref{eq:gravitino_in_thermal_bath}).
Because there is no dilution effect,
the abundance of this gravitino is decreased by a factor $B_{\mathrm{rad}}^{3/2}$
from the estimation eq.\,(\ref{eq:gravitino-refer}).
Thus, the constraint from the gravitino problem requires $B_{\mathrm{rad}}^{3/2}T_R<T_g$.

Finally, it should be noted that the gravitino problem in cases (ii) and (iii) may be more stringent,
if inflaton decay channels into heavier right-handed (s)neutrinos $N_2$ and/or $N_3$ are allowed
and branching ratio into all right-handed (s)neutrinos are almost the same order.
If $N_2$ and/or $N_3$ decay faster than $N_1$, their energy is converted into radiation
and the ratio between energy in $N_1$ and radiation after the decay of heavier (s)neutrinos is
at least $\mathcal{O}(1)$.
In these cases, gravitino produced by thermal scattering
after decay of heavier right-handed (s)neutrinos may result in serious constraint.

\subsection{Result for leptogenesis from inflaton decay}

\begin{figure}[!tp]
		\begin{center}			
			\includegraphics[width=.75\linewidth]{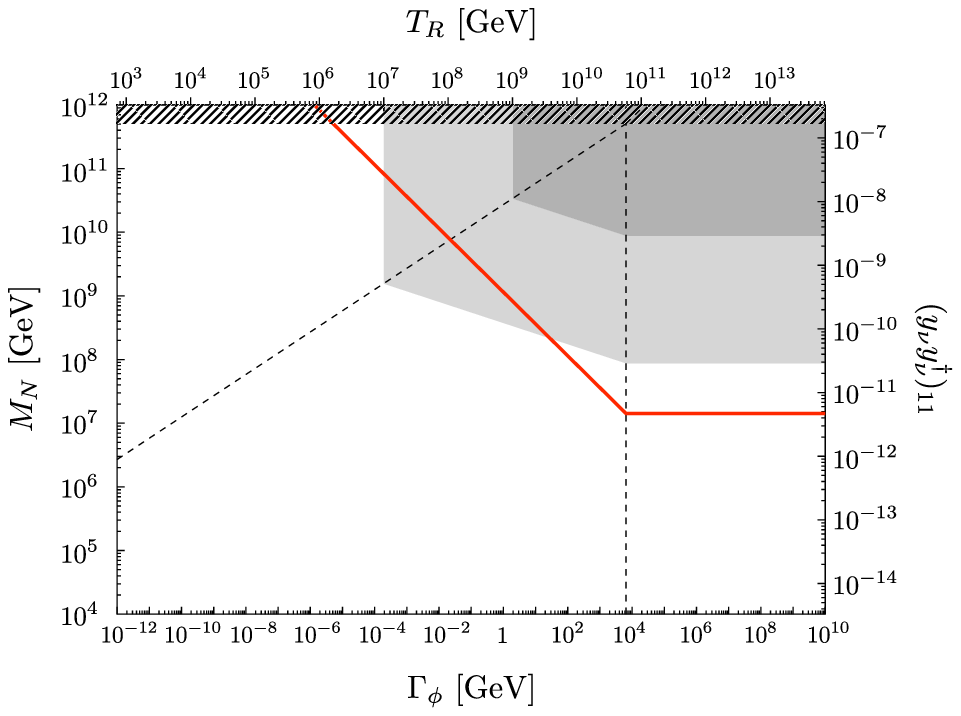}\\
			{\small(a) $\tilde{m}=10^{-5}{\mathrm{eV}},~m_\phi=10^{12}{\mathrm{GeV}}$}\\
		\end{center}
		\begin{center}			
			\includegraphics[width=.75\linewidth]{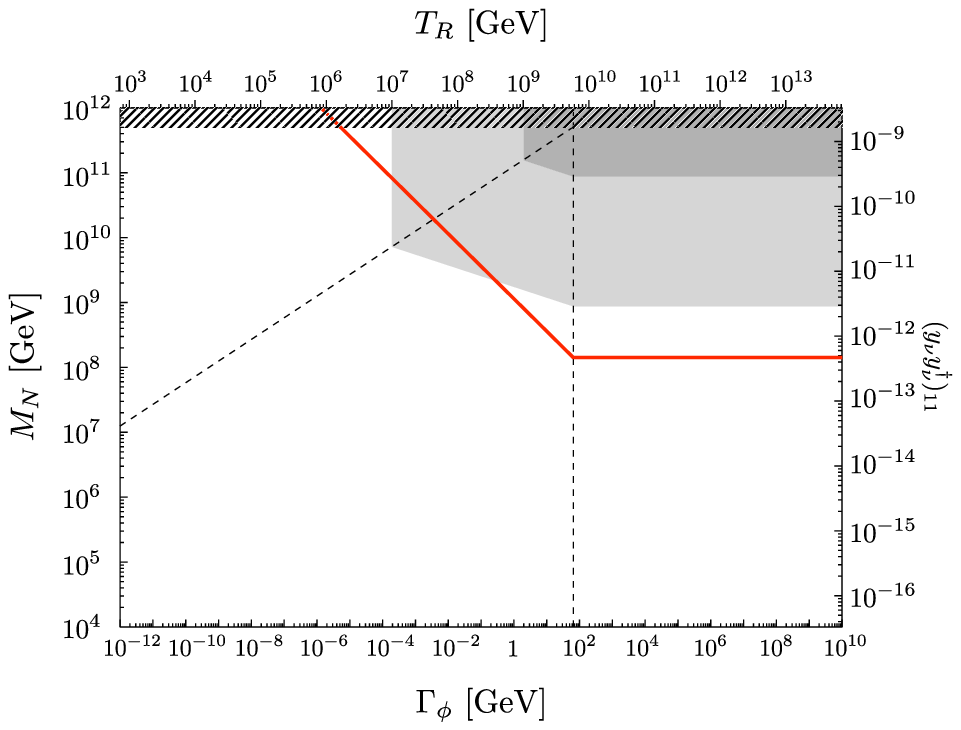}\\
			{\small(b) $\tilde{m}=10^{-7}{\mathrm{eV}},~m_\phi=10^{12}{\mathrm{GeV}}$}\\
		\end{center}
\end{figure}
\begin{figure}[!tp]	
		\begin{center}			
			\includegraphics[width=.75\linewidth]{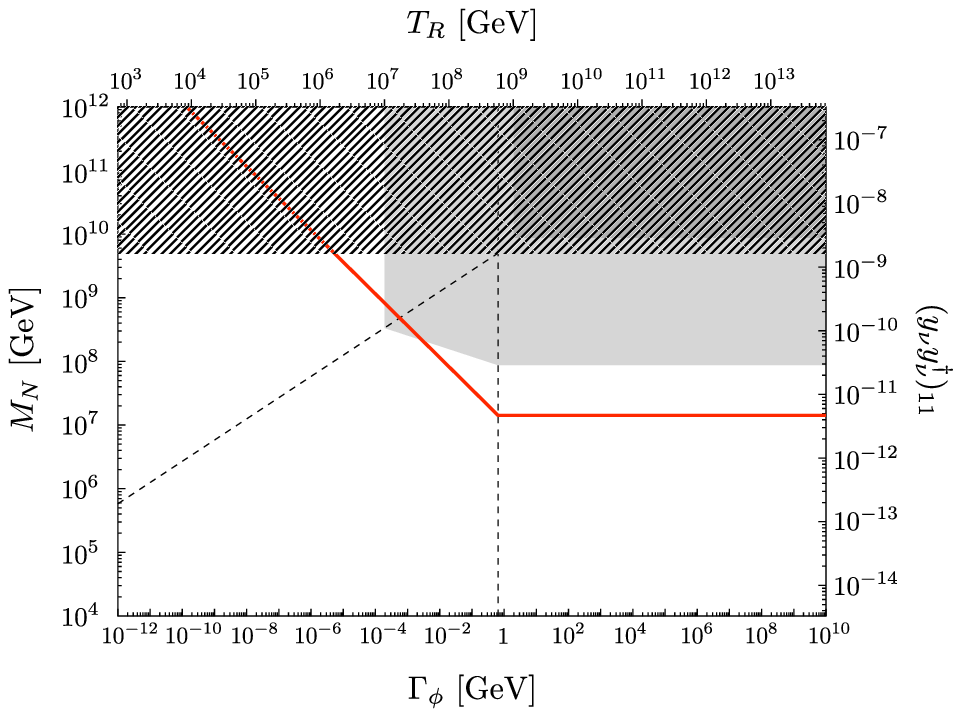}\\
			{\small(c) $\tilde{m}=10^{-5}{\mathrm{eV}},~m_\phi=10^{10}{\mathrm{GeV}}$}
		\end{center}
		\begin{center}			
			\includegraphics[width=.75\linewidth]{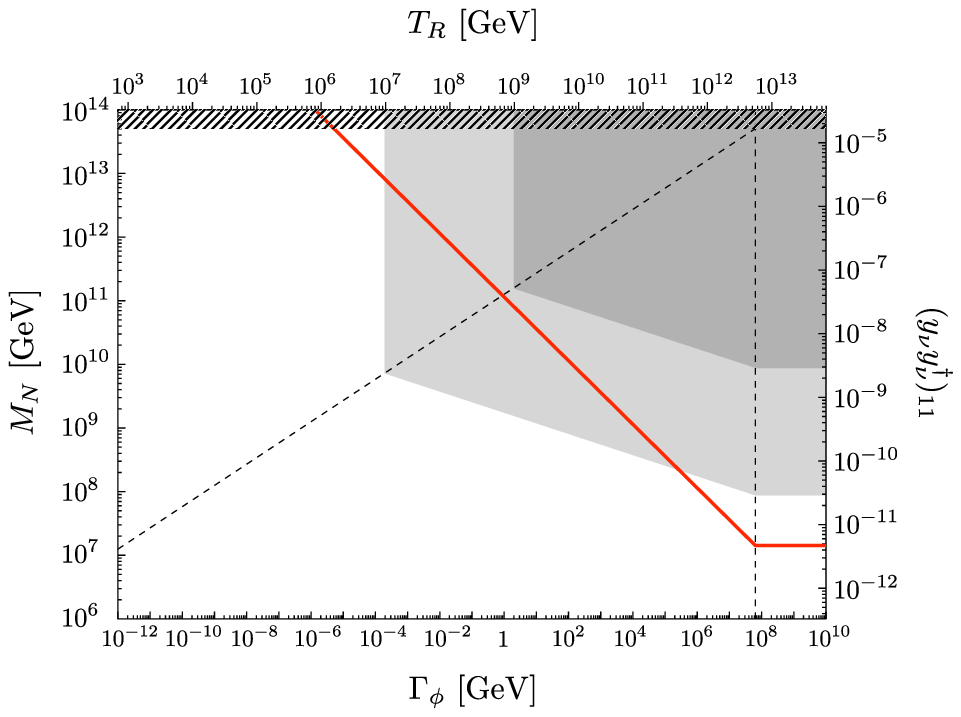}\\
			{\small(d) $\tilde{m}=10^{-5}{\mathrm{eV}},~m_\phi=10^{14}{\mathrm{GeV}}$}
		\end{center}
	\caption{\small 
	\label{fig:IDLG}
		We show our result in the $\Gamma_\phi-M_N$ plane
		for various choices of $\tilde{m}$ and $m_\phi$.
		On the solid (red) line, successful leptogenesis are realized.
		Values of $T_R$ and $(y_\nu y_\nu^\dagger)_{11}$ are also shown 
		on upper and right axes, respectively.
		The hatched region are excluded by the constraint $M_N < m_\phi/2$.
		Shaded region are excluded by the gravitino problem in cases 
		the constraint is $T_g=10^{7}{\mathrm{GeV}}$ (light) and $T_g=10^9{\mathrm{GeV}}$ (dark). 
		Two dashed lines distinguish between three cases:
		(i) in the upper-left region, (ii) in the right region, and (iii) in the middle region.
	}
\end{figure}

In fig.\,\ref{fig:IDLG}, we show our result in the $\Gamma_\phi-M_N$ plane.
We took $\delta_{\mathrm{eff}}=1$.
$B_{\mathrm{rad}} \lesssim 10^{-3}$ is assumed, and hence, $B_N\simeq1$.
We also show $T_R$ and $(y_\nu y_\nu^\dagger)_{11}$ on upper and right axes.
Four figures (a)-(d) are results for various choices of $\tilde{m}$ and $m_\phi$:
(a) $\tilde{m}=10^{-5}{\mathrm{eV}}$ and $m_\phi=10^{12}{\mathrm{GeV}}$,
(b) $\tilde{m}=10^{-7}{\mathrm{eV}}$ and $m_\phi=10^{12}{\mathrm{GeV}}$,
(c) $\tilde{m}=10^{-5}{\mathrm{eV}}$ and $m_\phi=10^{10}{\mathrm{GeV}}$, and
(d) $\tilde{m}=10^{-5}{\mathrm{eV}}$ and $m_\phi=10^{14}{\mathrm{GeV}}$.
The hatched region is excluded by the constraint $M_N < m_\phi/2$.
Two dashed lines show boundaries between the cases (i), (ii) and (iii),
which are upper left, right, and middle, respectively.
On the solid line (red), the abundance of the resultant 
baryon asymmetry of the universe is the best-fit value, $n_B/s=8.74\times10^{-11}$~\cite{Komatsu:2008hk}.
Above this line, the baryon asymmetry is larger than that of the universe.
Shaded regions are excluded by the gravitino problem for 
the constraint $T_g=10^{7}{\mathrm{GeV}}$ (light) and $T_g=10^9{\mathrm{GeV}}$ (dark).
Note that in fig.\,\ref{fig:IDLG} (c) the constraint from the gravitino problem for
$T_g=10^9{\mathrm{GeV}}$ disappears,
since the whole excluded region is included in the region $M_N > m_\phi/2$.

For low $T_R$ and large $M_N$, the case (i) is realized.
In this case, large $M_N$ is required for the successful leptogenesis,
in order to give sufficiently large $\epsilon$.
If the constraint on $T_g$ is as stringent as $T_g\sim 10^6{\mathrm{GeV}}$,
$M_N$ is required to be close to $m_\phi$.
For higher $T_R$, the case (ii) is realized, and the best-fit value of baryon asymmetry 
requires smaller $M_N$, i.e., small $y_\nu$.
The gravitino problem can be avoided if $T_N<T_g$ is satisfied.
As seen in eq.\,(\ref{eq:L_IDLG2}), the resultant baryon asymmetry is determined by properties of the neutrino sector.
The successful leptogenesis requires $T_N\gtrsim 10^6{\mathrm{GeV}}$.
The case (iii) is the intermediate region of these two regions.
In this case, larger $M_N$ than that in the case (ii) gives the best-fit value of baryon asymmetry.

\begin{figure}[!tp]
		\begin{center}			
			\includegraphics[width=.75\linewidth]{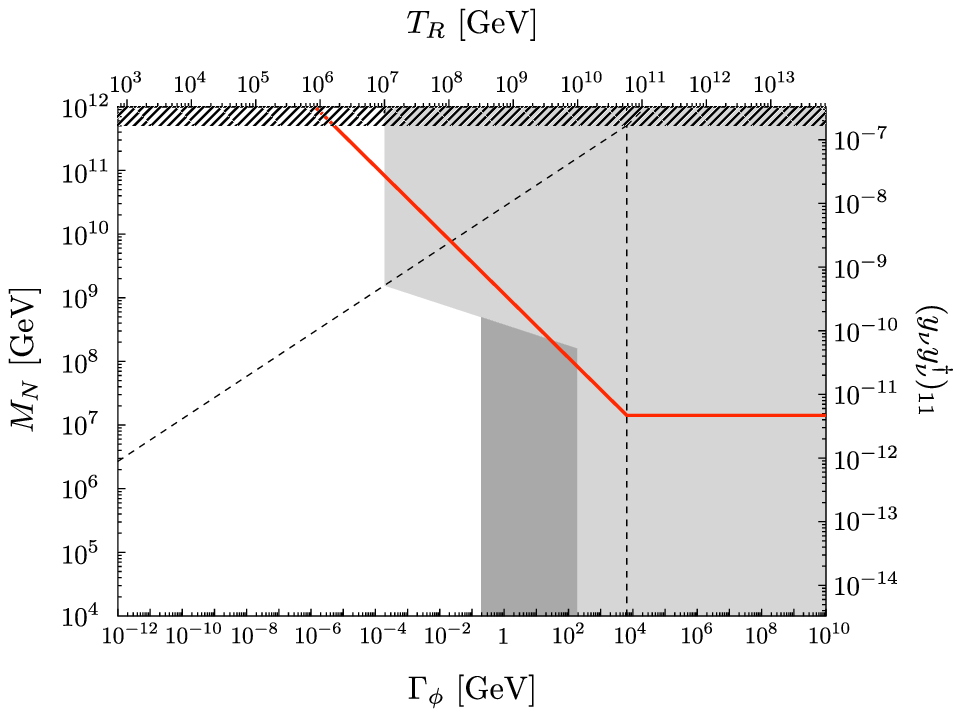}\\
			{\small(a) $\tilde{m}=10^{-5}{\mathrm{eV}},~m_\phi=10^{12}{\mathrm{GeV}}$}\\
		\end{center}
		\begin{center}			
			\includegraphics[width=.75\linewidth]{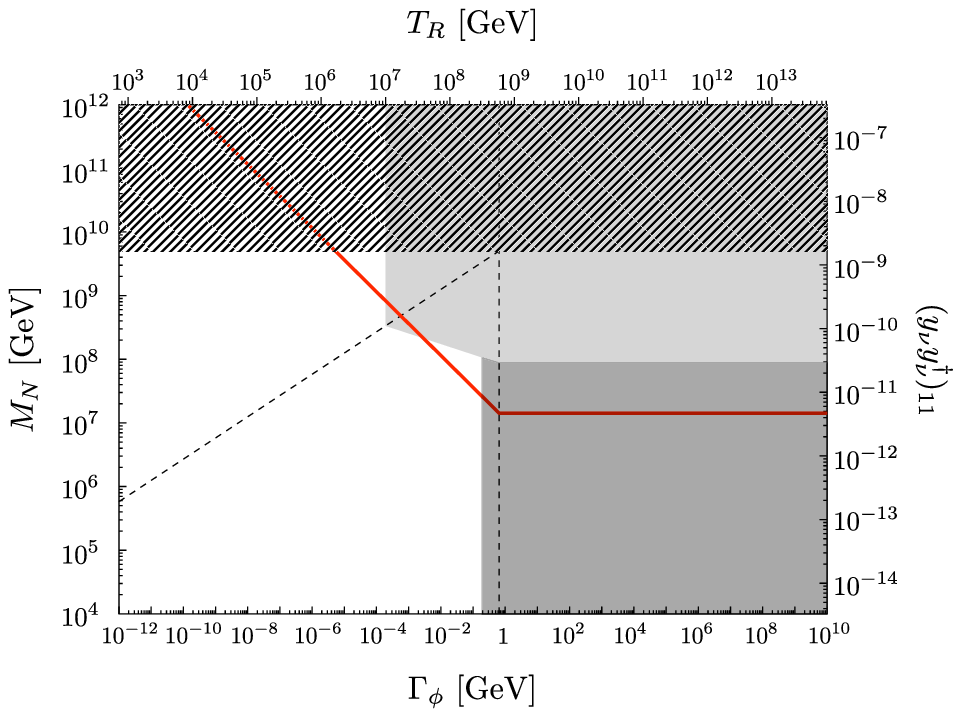}\\
			{\small(b) $\tilde{m}=10^{-5}{\mathrm{eV}},~m_\phi=10^{10}{\mathrm{GeV}}$}\\
		\end{center}
	\caption{\small 
	\label{fig:IDLG-2}
		Constraints from gravitino produced after the decay of inflaton.
		For (a) and (b), the same parameters as fig.\,\ref{fig:IDLG} (a) and (c), respectively,
		are chosen except for $B_{\mathrm{rad}}$.
		Here, we set $T_g=10^7{\mathrm{GeV}}$.
		The light shaded region is excluded for $B_{\mathrm{rad}}\ge10^{-2}$.
		For $B_{\mathrm{rad}}\ge10^{-1}$, 
		both light and dark shaded regions are excluded.
	}
\end{figure}

If $B_{\mathrm{rad}}$ is taken to be larger, the gravitino problem excludes high $T_R$ regions, 
as shown in figs.\,\ref{fig:IDLG-2} (a) and (b).
In these two figures, parameter regions for successful leptogenesis,
borders between three cases, and regions excluded by constraints are shown
in the similar way to fig.\,\ref{fig:IDLG}.
For figs.\,\ref{fig:IDLG-2} (a) and (b), the same parameters as figs.\,\ref{fig:IDLG} (a) and (c),
respectively,  are chosen except for $B_{\mathrm{rad}}$.
For simplicity, $B_N$ is still approximated to be $B_N=1$.
Light shaded region is excluded for $B_{\mathrm{rad}}\ge10^{-2}$,
while both light and dark shaded regions are excluded for $B_{\mathrm{rad}}\ge10^{-1}$.

For both cases (i) and (ii), the sufficient amount of lepton asymmetry requires
$T_{\mathrm{reh}} \gtrsim 10^6{\mathrm{GeV}}\times\delta_{\mathrm{eff}}^{-1}$.
In the case (ii), this corresponds to $M_N\gtrsim 10^6{\mathrm{GeV}}\times\delta_{\mathrm{eff}}$,
for the estimation of the largest $\tilde{m}$ that can avoid washout process,
$\tilde{m}<7\times10^{-4}{\mathrm{eV}}$.
Cases (ii) and (iii) are additional parameter regions to the case (i),
which has been discussed in many literatures.
Especially, case (iii) has not been seriously discussed

For example, the parameter region for the case (iii) can be advantageous in order to 
relax the non-thermal gravitino problem \cite{gravitino_from_inflaton1,gravitino_from_inflaton-2}.
It is argued that mixing between inflaton and SUSY-breaking field and/or
superconformal anomaly result in significant branching ratio of inflaton into gravitino,
if inflaton has large vev after the inflation.
The decay rate of inflaton into gravitino is determined by the inflation model.
Therefore, the branching ratio of inflaton into gravitino can be suppressed
if the total decay rate of inflaton is large.
Thus, if sufficiently strong interaction between inflaton and right-handed (s)neutrino is introduced,
and if right-handed (s)neutrino decay after the inflaton decay,
both non-thermal and thermal gravitino problems can be avoided,
while the baryon asymmetry of the universe can be explained simultaneously.

For inflation models in which inflaton has large vev after the inflation,
it may be difficult to explain small $M_N$ and large $\Gamma_\phi$
suitable for case (ii) shown in fig.\,\ref{fig:IDLG}.
Large decay rate $\Gamma_\phi$ requires strong coupling between 
inflaton and right-handed (s)neutrino,
while it means too large mass of right-handed (s)neutrino.
Let us consider hybrid-inflation model with the vev of inflaton (waterfall field)
$\langle \phi \rangle = M \sim 10^{15}{\mathrm{GeV}}$ and $m_\phi \sim 10^{12}{\mathrm{GeV}}$ as an example.
If an interaction
\begin{eqnarray}
	\label{eq:inflaton-RHneutrino}
	{\mathcal{L}}_{\phi-N}= \frac{1}{2}\frac{g_i}{M}\phi^2 N_iN_i,
\end{eqnarray}
is introduced, it can explain Majorana mass of right-handed neutrino and decay rate of inflaton,
as discussed in ref.\,\cite{Lazarides:1996dv}.
$\Gamma_\phi\sim10^4{\mathrm{GeV}}$ requires coupling constant
between inflaton and right-handed neutrino $g_1\gtrsim 10^{-4}$,
while this means mass of right-handed neutrino $M_N \gtrsim 10^{11}{\mathrm{GeV}}$,
which is far larger than $M_N\sim 10^7{\mathrm{GeV}}$ given in fig.\,\ref{fig:IDLG} (a).

For smaller $\tilde{m}$, $M_N$ corresponding to observed amount of baryon asymmetry becomes larger.
In addition, if $\delta_{\mathrm{eff}}$ is significantly smaller than 1,
and/or if the ratio between baryon asymmetry produced by sphaleron process 
and initial lepton asymmetry is smaller than eq.\,(\ref{eq:observed_baryon}),
observed amount of baryon asymmetry can be produced for larger $M_N$.
Of course, in latter two cases,
there is a trade-off between large $M_N$ and the condition from thermal gravitino problem.
In regards to this problem, case (iii) can be more plausible,
since required $M_N$ and $\Gamma_\phi$ are larger and smaller, respectively.
As a demonstration how non-thermal gravitino problem can be relaxed, 
let us consider $g_1\sim 10^{-5}$, $M_N\sim 10^{10}{\mathrm{GeV}}$
and $\Gamma_\phi \sim 10{\mathrm{GeV}}$,
where the case (iii) can explain the baryon asymmetry of the universe
for $T_g\gtrsim 10^9{\mathrm{GeV}}$.
According to ref.\,\cite{gravitino_from_inflaton-2}, the amount of non-thermally produced gravitino is estimated to be\footnote{
Here we assumed the SUSY-breaking scale $\Lambda$ is $m_\phi > \Lambda$.
}
\begin{eqnarray}
	\label{eq:non-thermal}
	Y_{3/2}\sim 10^{-12}\times\left( \frac{T_\phi}{10^6{\mathrm{GeV}}} \right)^{-1},
\end{eqnarray}
for above choices of $M$ and $m_\phi$.
For $\Gamma_\phi \sim 10{\mathrm{GeV}}$, $T_\phi\sim 10^9{\mathrm{GeV}}$.
Thus the amount of non-thermally produced gravitino eq.\,(\ref{eq:non-thermal}) is cosmologically allowed
for $m_{3/2}\gtrsim 10{\mathrm{TeV}}$,
even if hadronic branching ratio of gravitino is significantly large.
This example is interesting since both thermal and non-thermal gravitino problem 
is relaxed in some models of hybrid inflation, which is strictly constrained by these problems \cite{gravitino_from_inflaton1,gravitino_from_inflaton-2}.
It should be noted that inflaton does not decay into heavier two right-handed (s)neutrinos
if $g_2,g_3 \gtrsim 10^{-3}$, which is not so strongly hierarchical compared with $g_1\sim10^{-5}$.

The case (iii) is also interesting for $\tilde{m}\gtrsim 10^{-3}$.
As we discussed in eq.\,(\ref{eq:thermalize-case3}),
the condition for negligible washout is relaxed from eq.\,(\ref{eq:washout}),
since decay of right-handed (s)neutrino is delayed.
Therefore, the case (iii) is a possibility of leptogenesis with $\tilde{m}\gtrsim 10^{-3}$,
in addition to the case (i), where washout is avoided even for larger $\tilde{m}$.
It should be noted that the estimation eq.\,(\ref{eq:thermalize-case3}) is only an approximation.
Numerical calculation is necessary to determine the allowed range of $\tilde{m}$.

\section{Summary}

In this paper, we have investigated the leptogenesis scenario from
right-handed (s)neutrino produced by the inflaton decay,
surveying allowed parameter region systematically within a general framework. 
We have shown that the successful leptogenesis is possible both in $\Gamma_N>\Gamma_\phi$ 
and $\Gamma_\phi>\Gamma_N$ cases.
Especially, we considered all three cases including decay of relativistic right-handed (s)neutrino
and clarified the distinction of them.
The region where the lepton asymmetry is successfully generated
is given by $M_N\gtrsim 10^7{\mathrm{GeV}}$ and $T_R\gtrsim 10^6{\mathrm{GeV}}$.
If the constraint from the gravitino problem is more stringent than $T_g\gtrsim 10^6{\mathrm{GeV}}$,
this non-thermal leptogenesis scenario cannot explain the origin of baryon asymmetry.
We have also surveyed systematically the dependence of resultant lepton asymmetry
and the constraint from gravitino overproduction on relevant parameters,
and shown it in a brief form.
In the $\Gamma_N>\Gamma_\phi$ case, which has been considered in many literatures,
right-handed (s)neutrino decay instantaneously after they are produced.
Gravitino problem can be avoided for low reheating temperature 
$T_R<T_g$.
The resultant lepton asymmetry is proportional to $(T_R/m_\phi)\times M_N$.
Provided with $T_g\sim10^6{\mathrm{GeV}}$, $M_N/m_\phi\sim1$ is necessary 
to explain the observed baryon asymmetry.
On the other hand, in the $\Gamma_\phi>\Gamma_N$ case, the universe is once dominated by right-handed (s)neutrino.
Especially, the case (iii), where right-handed (s)neutrino decay during they are relativistic
has not been discussed seriously in literatures.
Thermal gravitino problem is avoided since the universe is reheated only after the decay of right-handed (s)neutrino,
and it constrains $T_N$.
In addition, if right-handed (s)neutrino are non-relativistic at the decay,
other unwanted relics produced after the decay of the inflaton can be diluted.
In this case, the amount of resultant lepton asymmetry is proportional to $T_N$.
Hence, larger $M_N$ is required for smaller $\tilde{m}$.
These two cases (ii) and (iii) are attractive since, for example, both
thermal and non-thermal gravitino problem can be relaxed.
For the case (ii), it seems to be difficult to satisfy required large $\Gamma_\phi$ and small $M_N$ simultaneously.
In this regards, the case (iii) is more plausible.
It is interesting that non-thermal gravitino problem in some models of hybrid inflation can be relaxed in our example.
The case (iii) is also interesting because the condition for avoiding washout processes
is relaxed compared with the case (ii).
Leptogenesis for relatively large washout parameter $\tilde{m}\gtrsim 10^{-3}{\mathrm{eV}}$ is possible 
not only in the case (i) but also in the case (iii).

\section*{Acknowledgments}

TT is grateful to Hitoshi Murayama and Kazunori Nakayama for useful communications.
This work was partially supported by JSPS research fellowships (TT). 


\end{document}